\documentclass[12pt]{iopart}
\usepackage{iopams,psfig,graphicx}
\usepackage[dvips]{color}
\newcommand{\ket}[1]{\vert #1 \rangle} \newcommand{\bra}[1]{\langle #1 \vert}
\newcommand{\dket}[1]{\vert #1 \rangle\rangle} 
\newcommand{\dbra}[1]{\langle\langle #1 \vert}

\newcommand{\dbraket}[2]{\langle\langle #1 \vert #2 \rangle\rangle}

\begin{document}
\title[Binary communication in single-mode and entangled noisy 
channels]{Binary optical communication in single-mode and entangled 
quantum noisy channels}
\author{Stefano Olivares\dag\ and Matteo G A Paris\ddag}
\address{\dag\ Dipartimento di Fisica and Unit\`a INFM, Universit\`a
degli Studi di Milano, via Celoria 16, I-20133 Milano, Italia}
\address{\ddag\ INFM Unit\`a di Pavia, via Bassi 6, I-27100, Pavia, Italia}
\begin{abstract}
We address binary optical communication in single-mode and entangled
quantum noisy channels. For single-mode we present a systematic comparison
between direct photodetection and homodyne detection in realistic
conditions, {\em i.e.} taking into account the noise that occurs both
during the propagation and the detection of the signals. We then consider
entangled channels based on twin-beam state of radiation, and show that
with realistic heterodyne detection the error probability at fixed channel
energy is reduced in comparison to the single-mode cases for a large range
of values of quantum efficiency and noise parameters.
\end{abstract} \date{\today}
\section{Introduction}
Classical information may be conveyed to a receiver through quantum
channels.  To this aim a transmitter prepares a quantum state drawn from a
collection of known states and sends it through a given quantum channel.
The receiver retrieves the information by measuring the channel, such to
discriminate among the set of possible preparations, and to determine the
transmitted signal. The encoding states are generally not orthogonal and
also when orthogonal signals are transmitted, they usually lose
orthogonality because of noisy propagation through the communication
channel.  Therefore, in general, no measurement allows the receiver to
distinguish perfectly between the signals \cite{Helart, Hel} and the need
of optimizing the detection strategy unavoidably arises. 
\par
In binary communication based on optical signals, information is encoded
into two quantum states of light. Amplitude modulation-keyed signals (AMK),
consist in two states of a single-mode radiation field, which are given by
$\varrho_j = |\psi_j\rangle\langle \psi_j|$, $j=1,2$, with $|\psi_1\rangle
= |\psi_0\rangle$ and $|\psi_2\rangle = D(\alpha)|\psi_0\rangle$,  where
$|\psi_0\rangle$ is a given {\em seed} state, usually taken as the vacuum,
$D(\alpha)=\exp(\alpha a^\dag - \alpha^* a)$ denotes the displacement
operator, and the complex amplitude $\alpha$ may be taken as real without
loss of generality.  The reason to choose AMK signals lies in the fact that
displacing a state is a simple operation, which is experimentally
achievable by a linear coupler and strong reference beam
\cite{displa}. Another binary encoding, based on displacement, is given by
phase shift-keyed signals (PSK) $|\psi_1\rangle = D(-\alpha)|\psi_0
\rangle$ and $|\psi_2\rangle=D(\alpha) |\psi_0\rangle$. In the following,
we will refer to the AMK situation only, though all the results hold also
for PSK.  The price to pay for such a convenient encoding stage is that,
for any choice of the seed state, the two signals are always not orthogonal,
and thus a nonzero probability of error appears in their discrimination
\cite{uir}.  If we consider equal {\em a priori} probabilities for the two
signals, the optimal quantum measurement to discriminate them
with minimum error probability is the projection-valued measure
$\{M_j\}_{j=1,2}$, $M_1 + M_2 = \mathbb{I}$, corresponding to \cite{Helart}
\begin{eqnarray}
M_j = \sum_k T[(-)^j \lambda_k] |\lambda_k\rangle\langle\lambda_k|\,,
\label{optPVM}\;
\end{eqnarray}
where $|\lambda_k\rangle$ is an eigenstate of the hermitian operator
$\Lambda=\rho_2-\rho_1$ with eigenvalue $\lambda_k$, and $T[x]$ is the unit
step function, which is zero for negative $x$, one for positive and $T(0) =
\frac12$. The probability of inferring the symbol $j$ when $i$ is
transmitted is given by $P(j|i)=\hbox{Tr}\{\varrho_i\:M_j\}$, $j,i=1,2$, such
that the average error probability in binary communication
is given by $P_{{\rm e}}=\frac12 \left\{P(1|2)+P(2|1)\right\}$.  
The minimum of $P_{\rm e}$, corresponding to the optimal measurement (\ref{optPVM}), 
reads as follows \begin{eqnarray}
\label{optPe}
P_{{\rm e}}=\frac12 \left(1-\sqrt{1-|\langle\psi_2|\psi_1\rangle|^2}
\right)\,,
\end{eqnarray}
and is known as the Helstrom bound \cite{Hel}. In particular, for a 
pair of AMK signals we have 
\begin{eqnarray}
\label{P:dolinar}
P_{{\rm e}}= \frac{1-\sqrt{1-\exp(-2N)}}{2}\,,
\end{eqnarray}
where with $N$ we denote the average number of photons in the channel
\emph{per use}, {\em i.e.} $N=\frac12 {\rm Tr}[a^{\dag}a(\rho_0 +
\rho_{\alpha})] =\frac12 |\alpha|^2 + n_0$, where $n_0=\langle
\psi_0|a^\dag a | \psi_0\rangle$ is the average photon number of the seed
state. For the sake of brevity, we will refer to $N$ also as to the {\em
energy of the channel}.  Notice that for a pair of PSK signals the same
bound in equation (\ref{P:dolinar}) holds; however, the expression for $N$
is now given by $N = |\alpha|^2 + n_0$.
\par
Binary communication has been the subject of much attention, mostly
concerning the design and the implementation of optimal quantum detection
processes, to distinguish nonorthogonal signals with reduced error
probability, possibly approaching the Helstrom bound given in equation
(\ref{optPe}). The relevant parameter in this optimization is the energy of
the channel, which itself limits the communication {\em rate} of the
channel.  After the pioneering work of Helstrom \cite{Helart}, a
near-optimum receiver for AMK signals based on direct detection 
was proposed in \cite{Ken}, whereas an optimum receiver approaching 
the minimum error probability (\ref{P:dolinar}) (based on photon counting 
and feedback) has been suggested in \cite{Dol}. More recently, various efforts
has been made to find out optimum detection operators and decision
processes for more general signals and in presence of noise \cite{Hir1,
Eld, Sasa, Enk}. Indeed, when one has at disposal a given set of quantum
signals, the problem becomes that of finding the optimal receivers
\cite{Ken, Dol} and detection schemes \cite{Sasa2, Miz} and to compare
their performances with those of realistic detectors. Following this way,
some studies were made on the effects of thermal noise on the optimum 
detection for a coherent AMK channel \cite{Hir5}. 
\par 
In this paper we focus our attention on protocols for binary communications
where both AMK signals and receivers can be realized with current
technology. As we will see, the various sources of loss and noise can be 
described as an overall Gaussian noise. Our analysis allows to unravel the 
different contributions and to compare receivers in realistic working regimes.
The purpose is twofold: on one hand we perform a systematic
comparison between direct and homodyne detection in presence of noise
during the propagation and the detection stages, in order to find in which
working regimes a receiver should be preferred. On the other hand, we
show that binary communication can be improved by using achievable sources
of entanglement and realistic heterodyne receivers.  Indeed, it has been
recently shown that in ideal conditions (perfect detection and noiseless
propagation) entanglement improves the performances of a binary channel,
{\em i.e.} it reduces the error probability in the discrimination of the
symbols \cite{par,ban}. Motivated by these results, we investigate the error
probability of entangled channels in realistic conditions, taking into
account the unavoidably noise that occurs during the propagation and the
detection.  Since we are interested in assessing entanglement as an
effective resource, we compare entangled channels with the corresponding
realistic single-mode channels.
\par
The paper is structured as follows. In Section \ref{Sec:Single:Mode} we
address single-mode channels that uses direct or homodyne detection as
receivers, and compare the corresponding error probabilities both in 
ideal and realistic situations, {\em i.e.} in presence of noise. In
section \ref{Sec:Ent:Modes} we describe a binary communication scheme based
on entangled twin-beam state of radiation that employs multiport homodyne
or heterodyne detection in the measurements stage. As we will see, there are 
regimes where the error probability is less than in a single mode channel, 
also when the noise affects propagation and detection. Finally, in section 
\ref{Sec:Concl} we summarize our results  giving some concluding remarks.
\section{Single Mode Communication}\label{Sec:Single:Mode}
\subsection{Direct detection - Ideal case}
\begin{figure}[h]
\begin{center}
\includegraphics[width=0.5\textwidth]{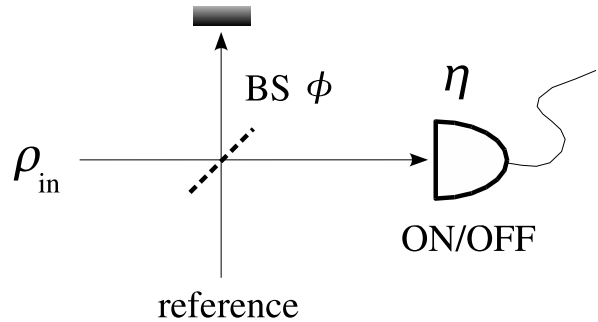}
\end{center}
\caption{Scheme of the receiver based on direct detection.
The signal to be processed is mixed 
at a beam splitter of transmissivity $\tau=\cos^2\phi$ with a given coherent
reference state. Then, at the output, one of the two modes is ignored or
absorbed, whereas the other one is revealed by {\sc on/off} photodetection.
In the ideal case $\rho_{\rm in}=\rho_\mu=|\mu\rangle\langle \mu|$, $\mu=0,
\alpha$, and the the detection efficiency is $\eta=1$. In presence
of noise, $\rho_{\rm in}$ is given by equation
(\protect\ref{matrix:therm:disp}) and $\eta<1$.} \label{f:ken}
\end{figure}

A scheme based on direct detection, to discriminate the set $\{|0\rangle,
|\alpha\rangle\}$, can be implemented as in figure \ref{f:ken}
\cite{Ken}. It  consists of a beam splitter (BS), in which the state to be
processed, either $\ket{0}$ or $\ket{\alpha}$, is mixed with a given
coherent reference state, say $\ket{\beta}$. The outgoing mode is
subsequently revealed by {\sc on/off} photodetection, {\em i.e.} by a
detector which checks the presence or absence or any number of photons.
\par
The operator describing the action of a BS on the modes $a$ and $b$ of the
field is
\begin{equation}\label{u:bs}
  U_{\phi}=\exp\left\{\phi (a^{\dag} b - a b^{\dag})\right\}\; ,
  \qquad \phi=\arctan \sqrt{\frac{1-\tau}{\tau}} \;,
\end{equation}
where $\tau = \cos^2 \phi$ is the BS transmissivity and $a$, $a^{\dag}$ and
$b$, $b^{\dag}$ are the annihilation and creation operators for the two
modes, respectively. If the input state is $\ket{\alpha}_a \ket{\beta}_b
\equiv \ket{\alpha}_a \otimes \ket{\beta}_b$, $\ket{\alpha}_a$ and
$\ket{\beta}_b$ being coherent states, the output state is given by
$\ket{\alpha}_a \ket{\beta}_b \longrightarrow \ket{\alpha\cos\phi +
\beta\sin\phi}_a\, \ket{-\alpha\sin\phi + \beta\cos\phi}_b$.
By choosing the amplitude of the reference as $\beta=-\alpha/\tan \phi$, we
obtain, at the output, the vacuum when the input state is $\ket{\alpha}$,
and $\ket{-\alpha \cos\phi}$ for vacuum input, in formula
\begin{eqnarray}
\ket{\alpha} \rightsquigarrow \ket{0}\qquad \mbox{and} \qquad
\ket{0} \rightsquigarrow \ket{-\alpha \cos\phi}.
\end{eqnarray}
In order to discriminate the two input signals, one performs a simple {\sc
on/off} photodetection: when the output is the vacuum the detector doesn't
click, otherwise it clicks. This measurement is described by the
probability operator-valued measure (POVM) $\{ \Pi_0, \Pi_1 \}$, where
$\Pi_0 = \ket{0}\bra{0}$ and $\Pi_0+\Pi_1 = \mathbb{I}$, {\em i.e.} we assumed 
unit detector efficiency. The error probability,
$K_{{\rm e}}$, is defined as:
\begin{eqnarray}\label{P:error}
K_{{\rm e}}=\frac12 \{ K(0|\alpha)+K(\alpha|0) \}\,,
\end{eqnarray}
where $K(0|\alpha)$ and $K(\alpha|0)$ are the probabilities of inferring that
the input state is $\ket{0}$ when it is actually $\ket{\alpha}$ and vice
versa. In our case
\begin{eqnarray}
K(0|\alpha) &=& \Tr\{  U_{\phi}\, \rho_{\alpha}\otimes\rho_{\beta}\,
U_{\phi}^{\dag} \, \Pi_1 \otimes \mathbb{I} \}\,, \\
K(\alpha|0) &=& \Tr\{  U_{\phi}\, \rho_{0}\otimes\rho_{\beta}\,
U_{\phi}^{\dag} \, \Pi_0 \otimes \mathbb{I} \}\,,
\label{Kal0}
\end{eqnarray}
where $\rho_{\mu}=\ket{\mu}\bra{\mu}$, $\mu = 0, \alpha, \beta$.
We obtain
\begin{eqnarray}
K(0|\alpha)&=&|\bra{0}\Pi_{1}\ket{0}|^2=0 \\ 
K(\alpha|0)&=&|\bra{-\alpha \cos\phi}\Pi_{0}\ket{-\alpha \cos\phi}|^2 =
\exp\{- |\alpha|^2 \cos^2\phi \},
\end{eqnarray}
such that Eq. (\ref{P:error}) rewrites as 
\begin{equation}\label{P:ken}
K_{{\rm e}}=\frac{\exp(-|\alpha|^2 \cos^2\phi)}{2} =
\frac{\exp(-2 N \cos^2\phi)}{2}.
\end{equation}
with $N = \Tr\{ a^{\dag}a\, (\rho_{0}+\rho_{\alpha}) \} = \frac12
|\alpha|^2$. Notice that in the limit $|\alpha|^2 \gg 1$ (relevant for
\emph{classical} communication) and $\cos^2\phi \rightarrow 1$, we have
$K_{{\rm e}} \rightarrow 2 P_{{\rm e}}$. This is usually summarized by saying 
that the measurement is asymptotically near optimal.
\subsection{Direct detection - Noise in propagation and detection}
\label{s:ken-noise}
In this section we take into account the effects due to the noise that
occurs in the propagation and the detection of the signals.  We model the
propagation in a noisy channel as the interaction of the single-mode
carrying the information with a thermal bath of oscillators at temperature
$T$. The dynamics is described by the Master equation
\begin{eqnarray}
\frac{{\rm d}\rho_t}{{\rm d}t} = \left\{
\Gamma (1+M) L[a]+\Gamma M L[a^{\dag}] \right\}\rho_t \, ,
\label{mast:eq:diss:one:mode}
\end{eqnarray}
where $\rho_t\equiv\rho(t)$ is the density matrix of the system at the time
$t$, $\Gamma$ is the damping rate, $M=({\rm e}^{\hbar
\omega/k_{B}T}-1)^{-1}$ is the number of thermal photons with frequency
$\omega$ at the temperature $T$, and $L[O]$ is the Lindblad superoperator,
$L[O]\rho_t=O\rho_t O^{\dag}-\frac12 O^{\dag}O\rho_t - \frac12 \rho_t
O^{\dag} O$. The term proportional to $L[a]$ describes the losses, whereas
the term proportional to $L[a^{\dag}]$ describes a linear phase-insensitive
amplification process. In other words, we are taking into account the
unavoidable dissipation and in-band amplifier noise. We are not considering
other sources of noise such as cross-talk and inter-symbol interference.
The Master equation (\ref{mast:eq:diss:one:mode}) can be transformed into a
Fokker-Planck equation for the Wigner function  
\begin{eqnarray}\fl
W_{\mu}(\zeta) \equiv W[\rho_{\mu}](\zeta) 
= \frac{1}{\pi^2} \int\! {\rm d}^2\lambda \,
{\rm e}^{\zeta\lambda^{*} - \zeta^{*}\lambda} \, {\rm Tr}\left\{ \rho_{\mu} \,
D(\lambda) \right\}
= \frac{2}{\pi} \, \exp\left\{ -2|\zeta - \mu|^2 \right\}\,,
\end{eqnarray}
with $\zeta\in \mathbb{C}$, $\rho_{\mu}=\ket{\mu}\bra{\mu}$,
$\mu=0,\alpha$, and $D(\lambda)$ is the displacement operator. Using the
differential representation of the Lindblad superoperator
\cite{nott,walls}, the Fokker-Planck equation associate to equation
(\ref{mast:eq:diss:one:mode}) is
\begin{equation}\label{f:p:single}\fl
\partial_t W_{\alpha,t}(x,y) = \left\{ \frac{\Gamma}{2}\left(
\partial_x x + \partial_y y \right) + \frac{\Gamma}{4}\left(M +
\frac12\right)\left( \partial_x^2 + \partial_y^2 \right)
\right\}W_{\alpha,t}(x,y)
\end{equation}
where we put $\zeta=x+iy$ and $W_{\mu,t}(\zeta)$ is the Wigner function of
the system at time $t$, for initial state $\rho_\mu$. The solution of
equation (\ref{f:p:single}) can be written as the convolution
\begin{equation}
W_{\mu,t}(x,y) = \int\!\!\int {\rm d}x' {\rm d}y'\: W_{\mu,0} (x', y')
\: G_{t}(x|x') \: G_{t}(y|y')\,,\label{sol}
\end{equation}
where the Green function $G_{t}(x_j|x'_j)$ is given by
\begin{equation}
G_{t}(x_j|x'_j)=\frac{\displaystyle 1}{\displaystyle \sqrt{2 \pi D^2}}
\exp\left\{ -\frac{\displaystyle (x_j - x'_j {\rm e}^{-\frac12
\Gamma t})^2}{\displaystyle 2 D^2} \right\}\,,\label{gre}
\end{equation}
with $D^2=\frac12 (M+\frac12)(1-{\rm e}^{-\Gamma t})$.
Using the equations (\ref{sol}) and (\ref{gre}), we arrive at
\begin{equation}\label{wigner:therm:disp}
W_{\mu,t}(\zeta) = \frac{1}{\pi \Delta_{M\Gamma}^2}\, \exp\left\{
-\frac{|\zeta - \mu\: {\rm e}^{-\frac12 \Gamma t}|^2}{\Delta_{M\Gamma}^2}
\right\},
\end{equation}
with $\Delta_{M \Gamma}^2=\frac12 [1+2M(1-{\rm e}^{-\Gamma t})]$. The Wigner
function in equation (\ref{wigner:therm:disp}) corresponds to the density
matrix of a displaced thermal state \cite{marian} 
\begin{equation}\label{matrix:therm:disp}
\rho_{\alpha}(t) \equiv \rho_{M'}=D(\alpha')\, \nu_{M'} \,D^{\dag}(\alpha')\,,
\end{equation}
where $\alpha' = \alpha\: {\rm e}^{-\frac12 \Gamma t}$ and $\nu_{M'}$ is a
thermal state
\begin{equation}\label{thermal:state}
\nu_{M'} = \frac{1}{1+M'} \left( \frac{M'}{1+M'} \right)^{a^{\dag}a}
\end{equation}
with $M'=M(1-{\rm e}^{-\Gamma t})$ average number photons.
\par 
Equation (\ref{matrix:therm:disp}) describes the signal arriving at the
receiver (figure \ref{f:ken}). Let us now consider the noise in the
detection stage.  At first we have to choose the reference state. Since
this receiver is based on the interference between the signal and the
reference, it turns out that, in presence of propagation noise, the optimal
reference is the coherent state $\ket{\beta'}$, $\beta' = -\alpha' /
\tan\phi$.  Moreover, if {\sc on/off} detection is not ideal, we must
consider the finite detector efficiency $\eta$. In this case, the
POVM describing the measurement is $\Pi_{0}(\eta) + \Pi_{1}(\eta) =
\mathbb{I}$, where
\begin{eqnarray}
\Pi_{0}(\eta)&=&\sum_{n=0}^{\infty}\,(1-\eta)^n \ket{n}\bra{n}.
\end{eqnarray}
In order to evaluate the detection probabilities we use the fact 
that the trace between two operators, $O_1$ and $O_2$, can be written as 
the phase-space integral
\cite{glau}
\begin{equation}\label{wigner:trace}
{\rm Tr}\{ O_1 O_2 \} \equiv \pi \int\! {\rm d}^2 \zeta\, W[O_1](\zeta)\,
W[O_2](\zeta)\,,
\end{equation}
where the Wigner function of a generic operator $O$ is defined as
\begin{equation}\label{wigner:operator}
W[O](\zeta)\equiv\frac{1}{\pi^2}\int {\rm d}^2 \lambda\, 
{\rm e}^{\zeta \lambda^* - \zeta^* \lambda}\, {\rm Tr}\{O\, D(\lambda)\}\,.
\end{equation}
The Wigner of the POVM element $\Pi_0$ is then
\begin{equation}
W[\Pi_0(\eta)](\zeta)=\frac{1}{\eta}\, \frac{1}{\pi
\Delta_{\eta}^2}\, \exp\left\{ -\frac{|\zeta|^2}{\Delta_{\eta}^2}
\right\}\,,
\end{equation}
with $\Delta_{\eta}^2 = (1-\eta)/(2\eta)$. 
\par
The expression (\ref{Kal0}) for probability $K_{\eta,\Gamma,M}(\alpha|0)$ to 
infer $\ket{\alpha}$ when $\ket{0}$ is sent is modified as follows
\begin{equation}\label{p:0:alpha:ken:noise}
K_{\eta,\Gamma,M}(\alpha|0) = \Tr\{ U_{\phi}\,
\nu_{M'}\otimes\ket{\beta'}\bra{\beta'} U^{\dag}_{\phi}\, \Pi_0(\eta)
\otimes \mathbb{I} \}\,.
\end{equation}
By equation (\ref{wigner:trace}) we have
\begin{equation}\label{p:0:alpha:ken:noise:wig1}\fl
K_{\eta,\Gamma,M}(\alpha|0) = \pi^2 \int\! {\rm d}^2 \gamma \, {\rm d}^2
\lambda\, W[U_{\phi}\, \nu_{M'}\otimes\ket{\beta'}\bra{\beta'}\,
U^{\dag}_{\phi}](\gamma,\lambda)\, W[\Pi_0(\eta)](\gamma)\,,
\end{equation}
which, using 
\begin{eqnarray}\fl
W[U_{\phi}\, \nu_{M'}\otimes\ket{\beta'}\bra{\beta'}\,
U^{\dag}_{\phi}](\gamma,\lambda) &=& W[\nu_{M'}](\gamma\cos\phi +
\lambda\sin\phi) \nonumber\\
&\mbox{}& \hspace{1cm} \times
W[\ket{\beta'}\bra{\beta'}](-\gamma\sin\phi+\lambda\cos\phi)\,,
\end{eqnarray}
leads to 
\begin{equation}
K_{\eta,\Gamma,M}(\alpha|0) = 
\frac{ \exp\left\{ -\frac{ 2 \eta N {\rm e}^{-\Gamma t} \cos^2\phi }{ 1 +
\eta M (1-{\rm e}^{-\Gamma t}) \cos^2\phi} \right\}}{1+ 
\eta M (1-{\rm e}^{-\Gamma t}) \cos^2\phi }\,.
\end{equation}
On the other hand, for the probability $K_{\eta,\Gamma,M}(0|\alpha)$ we have 
\begin{eqnarray}\fl
K_{\eta,\Gamma,M}(0|\alpha) &=& \Tr\{ U_{\phi}\,D(\alpha') \nu_{M'} 
D^{\dag}(\alpha') \otimes \ket{\beta'}\bra{\beta'} U^{\dag}_{\phi}\, \Pi_1(\eta) \otimes
\mathbb{I} \}\,,\nonumber\\
&=& 1 - \Tr\{ U_{\phi}\,D(\alpha')\, \nu_{M'}\, D^{\dag}(\alpha') \otimes
\ket{\beta'}\bra{\beta'} U^{\dag}_{\phi}\, \Pi_0(\eta) \otimes \mathbb{I}
\}\,,\nonumber\\
&=& \frac{\eta M (1-{\rm e}^{-\Gamma t}) \cos^2\phi}{1 + \eta M (1-{\rm
e}^{-\Gamma t})\cos^2\phi }\,, \label{p:alpha:o:ken:noise}
\end{eqnarray}
and, finally, the overall error probability in presence of 
noise is 
\begin{equation}\label{P:ken:eta}\fl
K_{{\rm e}}(N,\eta,\Gamma,M) = \frac{\eta M (1-{\rm e}^{-\Gamma t})
\cos^2\phi + \exp\left\{ -\frac{ 2 N \eta {\rm e}^{-\Gamma t} \cos^2\phi }{
1 + \eta M (1-{\rm e}^{-\Gamma t}) \cos^2\phi } \right\}}{2[1 + \eta M
(1-{\rm e}^{-\Gamma t})\cos^2\phi]}\,,
\end{equation}
which reduces to equation (\ref{P:ken}) in the limits $\eta\rightarrow 1$ and
$\Gamma, M\rightarrow 0$. Notice that the result of 
equation (\ref{P:ken:eta}) corresponds to the presence of an overall  
Gaussian noise with parameter $\sigma^2 = \eta M (1-{\rm e}^{-\Gamma t})$ 
plus dissipation 
of the signal ($\alpha\rightarrow \alpha\: {\rm e}^{-\frac12 \Gamma t}$). 
Our analysis allows to unravel the different contributions to $\sigma$.
\subsection{Homodyne detection - Ideal case} 
\begin{figure}[htb]
\begin{center}
\includegraphics[width=0.4\textwidth]{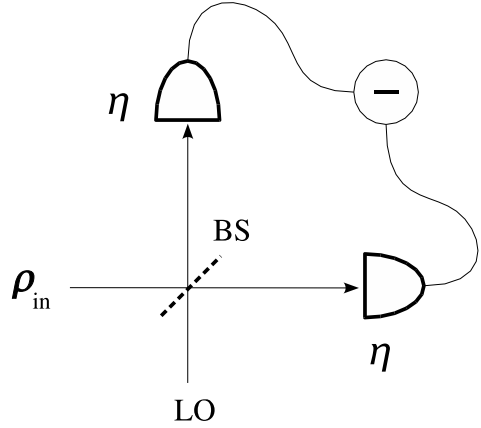}
\end{center}
\caption{Scheme of the homodyne receiver. The signal to be processed is mixed
at a balanced beam splitter with a highly excited coherent reference state,
usually referred to as the local oscillator (LO). Then, the difference
photocurrent is measured at the output. In the ideal case
$\rho_{\rm in}=\rho_\mu=|\mu\rangle \langle \mu|$, $\mu=0,\alpha$, and the the
detection efficiency is $\eta=1$.  In presence of noise, $\rho_{\rm in}$ is
given by equation (\protect\ref{matrix:therm:disp}) and $\eta<1$.}
\label{f:hom}
\end{figure}

An alternative receiver for single mode communication is provided by
homodyne detection, which offers the advantage of amplification from a
local oscillator, avoiding the need of single-photon avalanche
photodetectors \cite{yuenH}. A schematic diagram of the balanced homodyne
detection is shown in figure \ref{f:hom}: here the signal interferes with a
local oscillator (LO), {\em i.e.} a highly excited coherent state, in a
balanced BS [this corresponds to put $\phi = \pi/4$ in equation
(\ref{u:bs})].  After the BS the two modes are detected and the difference
photocurrent is electronically formed. For unit quantum efficiency of
photodiodes, the POVM of the detector is $\Pi_{x}=\ket{x}\bra{x}$, with
\begin{equation}
\ket{x}=\left( \frac{2}{\pi} \right)^{1/4} {\rm e}^{-x^2} \sum_{n=0}^{\infty}
\frac{H_n (\sqrt{2}x)}{\sqrt{n! \, 2^n}}\,\ket{n}\label{xquad}
\end{equation}
being an eigenstate of the quadrature operator $x=\frac12(a+a^\dag)$ of the
measured mode.  In equation (\ref{xquad}) $H_n(x)$ denotes the $n$-th
Hermite polynomials.  The probability density $p(x|\mu)$ of obtaining the
outcome $x$ from homodyne detection with input state $\ket{\mu}$,
$\mu=0,\alpha$, is
\begin{equation}\label{hom:prob:dens:alpha}
p(x|\mu)={\rm Tr}\{\ket{\mu}\bra{\mu} \, \Pi_{x} \} =
\sqrt{\frac{2}{\pi}} \, \exp\{ -2(x-\mu)^2 \}\:.
\end{equation}
In equation (\ref{hom:prob:dens:alpha}) $\alpha$ is assumed as real.
Equivalently, if $\alpha \in \mathbb{C}$, the same result may be obtained
by measuring a suitable quadrature $x_{\varphi}=\frac12(a^{\dag}
e^{\varphi}+a e^{-\varphi})$, with $\varphi={\rm arg}(\alpha)$.
The minimum error probability for the homodyne receiver is given by
\begin{eqnarray}
H_{{\rm e}}&=&\frac12 \left\{ H(0|\alpha)+H(\alpha|0) \right\}
\nonumber \\
\mbox{} &=& \frac12 \left\{ \int_{-\infty}^{\alpha/2} {\rm d}x \,
p(x|\alpha) + \int_{\alpha/2}^{+\infty} {\rm d}x \,
p(x|0)\right\} \nonumber \\
\mbox{} &=& \frac12 \left\{ 1-{\rm Erf}\left[\sqrt{N}\right]
\right\} \label{P:hom},
\end{eqnarray}
where $H(0|\alpha)$ and $H(\alpha|0)$ are the probabilities of inferring
the signal $\ket{0}$ when it is actually $\ket{\alpha}$ and vice versa.
${\rm Erf} [a]=\frac{2}{\sqrt{\pi}}\int_{0}^{a} {\rm d}\zeta \, {\rm
e}^{-\zeta^2}$ denotes the error function.  Notice that, in general, the
error probability depends on the choice of a threshold parameter $\Lambda$,
\emph{i.e.}
\begin{eqnarray}
H_{{\rm e}} (\Lambda)&=& \frac12 \left\{
\int_{-\infty}^{\Lambda} {\rm d}x \, p(x|\alpha) +
\int_{\Lambda}^{+\infty} {\rm d}x \, p(x|0)\right\}\nonumber\\ 
\mbox{} &=& \frac12 \left\{ 1 -\frac12 \left( {\rm Erf}[\sqrt{2}\Lambda] +
{\rm Erf}[\sqrt{2}(\alpha-\Lambda)] \right) \right\}\:. \label{infer:con}
\end{eqnarray}
In our case this probability is minimized when $\Lambda=\alpha/2$, thus
leading to the result in equation (\ref{P:hom}). In the limit $N\gg 1$,
equation (\ref{P:hom}) reduces to
\begin{equation}
H_{{\rm e}} \approx \frac{{\rm e}^{-N}}{2\sqrt{\pi N}}\,.
\end{equation}
Homodyne detection provides a better discrimination of the signals than
direct detection, {\em i.e.} $H_{\rm e}<K_{\rm e}$ if the energy of the
channel is below a threshold that monotonically increases as the
transmissivity of the BS in the receiver decreases.  In the limit of
$\tau\rightarrow 1$ in direct detection, we have $H_{\rm e}<K_{\rm e}$ for
$N\lesssim 0.77$ while, as an example, for $\tau=0.9$ we have $H_{\rm
e}<K_{\rm e}$ for $N\lesssim 1.10$. 
\subsection{Homodyne detection - Noise in propagation and detection}
As already shown in section \ref{s:ken-noise}, if noise affects the
propagation of the signal, the state arriving at the homodyne receiver is
no longer a pure state, and is given in equation (\ref{matrix:therm:disp}).
Moreover, when also homodyne detection is not ideal, the POVM of the
receiver is a Gaussian convolution of the ideal POVM
\begin{equation}\label{Pi:hom:eta}
\Pi_{x}(\eta) = \frac{1}{\sqrt{2 \pi \sigma_{\eta}^2}} \int {\rm d}y \,
\exp\left\{-\frac{(y-x)^2}{2\sigma_{\eta}^2}\right\} \, \Pi_{y}\, ,
\end{equation}
where $\sigma_{\eta}^2 = 
(1-\eta)/(4\eta)$, and $\eta$ is the quantum efficiency of both photodiodes 
involved in homodyning (we assume that they have the same quantum efficiency). 
The Wigner function of $\Pi_x(\eta)$ is given by 
\begin{eqnarray}
W[\Pi_x(\eta)](\zeta) &\equiv& W[\Pi_x(\eta)](\hbox{Re}[\zeta])\nonumber\\ 
&=& \frac{1}{\sqrt{2 \pi \sigma_{\eta}^{2}}}\,
\exp\left\{ -\frac{(\hbox{Re}[\zeta] - x)^2}{2 \sigma_{\eta}^{2}}
\right\}\,.
\end{eqnarray}
Taking into account all the sources of noise, the probability density  of
equation (\ref{hom:prob:dens:alpha}) becomes
\begin{eqnarray}
p_{\eta,\Gamma,M}(x|\mu)&=&{\rm Tr}\{D(\mu')\, \nu_{M'}\,
D^{\dag}(\mu') \, \Pi_{x}(\eta) \}\,,
\end{eqnarray}
with $\mu'=\mu\: {\rm e}^{-\frac12 \Gamma t}$, $\mu=0,\alpha$, and $\nu_{M'}$
given by equation (\ref{thermal:state}). In this way, using the Wigner
functions and thanks to equation (\ref{wigner:trace}), the error probability
reads as follows
\begin{equation}
H_{{\rm e}}(N,\eta,\Gamma,M) = \frac12 \left\{ 1-{\rm Erf}
\left[\frac{\sqrt{\eta N}\, {\rm e}^{-\frac12 \Gamma t}}{\sqrt{1 + 2\eta M
(1-{\rm e}^{-\Gamma t})}}
\right]\right\}\,,
\end{equation}
which, in the limit $\eta N\: {\rm e}^{- \Gamma t} \gg 1 +
2\eta M (1-{\rm e}^{-\Gamma t})$, reduces to
\begin{equation}\label{P:hom:eta:limit}
H_{{\rm e}}(N,\eta,\Gamma,M) \approx \frac{ \sqrt{1 + 2\eta M (1-{\rm
e}^{-\Gamma t})}\, \exp\left\{ -\frac{\eta N\: {\rm e}^{-\Gamma t}}{1+2\eta
M (1-{\rm e}^{-\Gamma t})} \right\}}{2 \sqrt{\pi \eta N}\: {\rm
e}^{-\frac12 \Gamma t}}\,.
\end{equation}
In the next section we compare $H_{\rm e}$ with the corresponding error
probability in direct detection. 
\subsection{Direct vs homodyne detection} 
\begin{figure}[htb]\begin{center}
\includegraphics[width=0.6\textwidth]{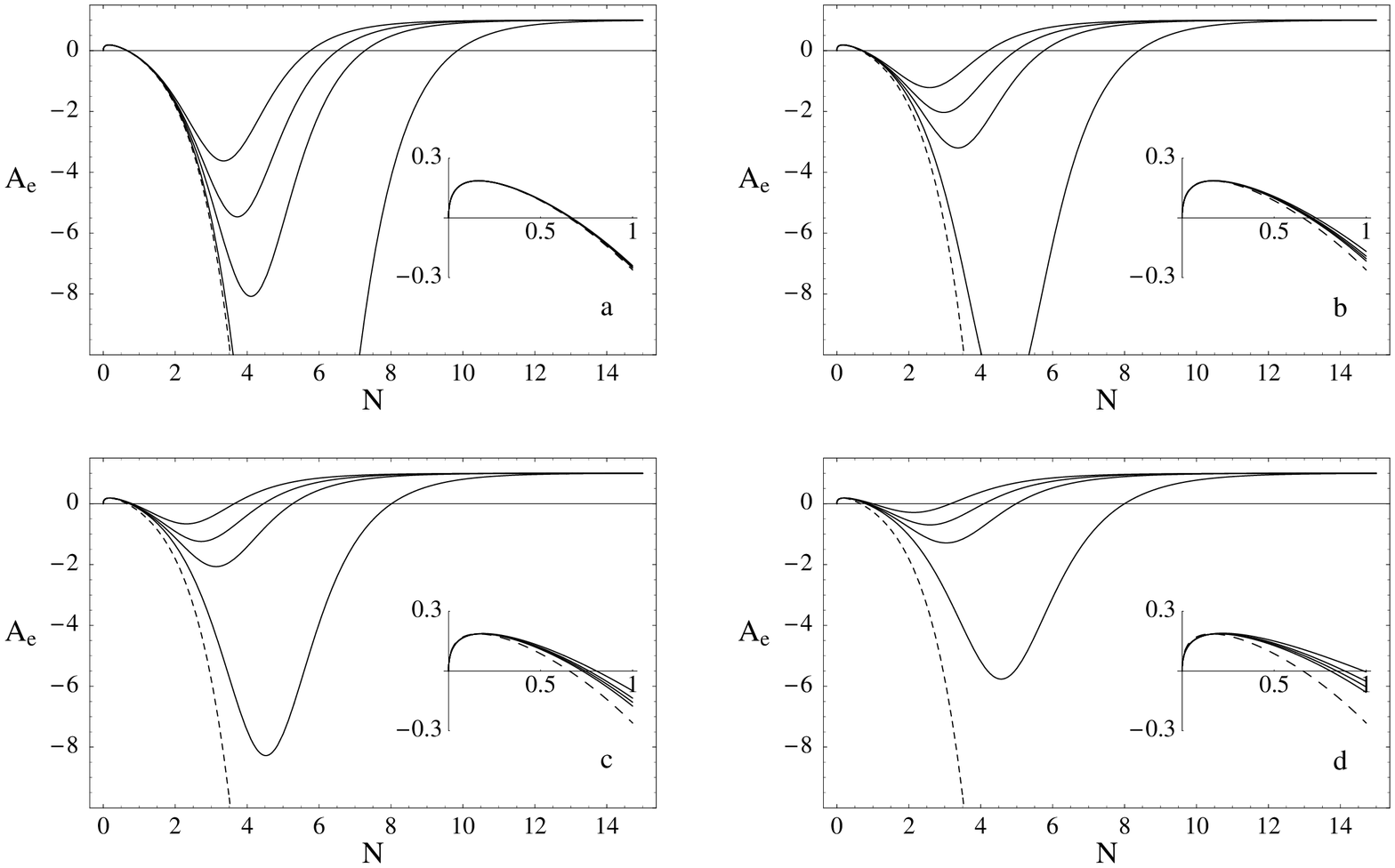}\end{center}
\caption{Plots of $A_{{\rm e}}(N,\eta_{{\rm ken}},\eta_{{\rm hom}},\Gamma,
M)$, defined in equation (\ref{A:e}), as a function of $N$ for different
values of $\Gamma t$ and $M$. The {\sc on/off} and homodyne detection
efficiencies are chosen to be the realistic values $\eta_{{\rm ken}}=0.95$
and $\eta_{{\rm hom}}=0.85$, respectively, and $\cos^2\phi=0.99$. The
dashed line is $A_{{\rm e}}$ with $\Gamma, M = 0$. In all the plots the
solid lines correspond to (from bottom to top) $M=5\, 10^{-3}, 5\, 10^{-2},
0.1$ and $0.2$, respectively, while $\Gamma t$ is: (a)  $\Gamma t=10^{-2}$,
(b) $\Gamma t=5\, 10^{-2}$, (c) $\Gamma t=0.1$, (d) $\Gamma t=0.2$. The
insets refer to the region $0<N<1$. When $A_{{\rm e}}>0$ one has $H_{{\rm
e}}<K_{{\rm e}}$.\label{f:kenvshom}}
\end{figure}

In order to individuate the working regimes where homodyne detection 
provides better performances, {\em i.e.} lower error probability than 
direct detection, we define the following quantity
\begin{equation}\label{A:e}
A_{{\rm e}}(N,\eta_{{\rm ken}},\eta_{{\rm hom}},\Gamma, M) = 1 -
\frac{H_{{\rm e}}(N,\eta_{{\rm hom}},\Gamma,M)}{K_{{\rm e}}(N,\eta_{{\rm
ken}},\Gamma,M)}\,,
\end{equation}
where $\eta_{{\rm ken}}$ and $\eta_{{\rm hom}}$ are the {\sc on/off} and
homodyne detection efficiencies, respectively. When $A_{{\rm e}}>0$
homodyne receiver's error probability is the lowest. In figure
\ref{f:kenvshom} we plot $A_{{\rm e}}$ as a function of the energy of the
channel $N$ for different values of the other parameters. 
For given values of $\eta_{{\rm ken}}$, $\eta_{{\rm hom}}$, $\Gamma$ and
$M$ we have two different thresholds for the energy channel, namely $N^{\rm
(hom)}_{{\rm th},j}(\eta_{{\rm ken}},\eta_{{\rm hom}},\Gamma,M)$, $j=1,2$,
such that $A_{{\rm e}}=0$.  For $N \le N^{\rm (hom)}_{{\rm th},1}$ we have
a small interval where homodyne detection should be preferred to direct
one; as $N$ increases, we find a window ($N^{\rm (hom)}_{{\rm th},1} < N
\le N^{\rm (hom)}_{{\rm th},2}$), where $H_{{\rme}} > K_{{\rm e}}$ and,
finally, a last region for $N > N^{\rm (hom)}_{{\rm th},2}$ where homodyne
detection returns to be definitely better (see table \ref{t:due}). This 
result is due to the presence of thermal noise (\emph{i.e.} $T \ne 0$). In
fact, as one can easily see from equations (\ref{P:ken:eta}) and
(\ref{P:hom:eta:limit}) one has, independently on $M$
\begin{eqnarray}
\lim_{N\rightarrow \infty} K_{{\rm e}}(N,\eta_{{\rm ken}},\Gamma,M) &=&
\frac{\eta_{{\rm ken}} M (1-{\rm e}^{-\Gamma t}) \cos^2\phi}{2[1 + \eta_{{\rm
ken}} M (1-{\rm e}^{-\Gamma t})\cos^2\phi]}\,, \\
\lim_{N\rightarrow \infty} H_{{\rm e}}(N,\eta_{{\rm hom}},\Gamma,M) &=&
0\,.
\end{eqnarray}
\par
In summary, homodyne detection provides better results for either 
small or large values of the channel energy $N$, whereas for intermediate
values of $N$ the optimal choice is represented by direct detection.
The width of this intermediate region decreases as the noise increases,
{\em i.e.} as the value of both $\Gamma$ and $M$ increases. We therefore 
conclude
that homodyne detection is a more robust receiver in presence of noise.  As
concern quantum efficiency, we have, as one may expect, that the
performances of each detector improves increasing the corresponding $\eta$. 
\section{Binary communication in entangled channels}\label{Sec:Ent:Modes} 
Entanglement is a key feature of quantum mechanics.  The quantum
nonlocality due to entanglement has been, in the last decade,  harnessed
for practical use in the quantum information technology
\cite{popescubook,nielsen}. Entanglement has become an essential resource
for quantum computing \cite{nielsen}, quantum teleportation \cite{kim},
dense coding \cite{ban}, and secure cryptographic protocols \cite{nielsen}
as well as for improving optical resolution \cite{fabre}, spectroscopy
\cite{spettr}, and general quantum measurements \cite{em}.  Here we show
how entanglement, and in particular entangled states that can be realized
by current optical technology, can be used to improve binary
communications, {\em i.e.} to reduce the error probability at fixed energy
of the channel.
\subsection{Heterodyne detection - Ideal case}
\begin{figure}[htb]
\begin{center}
\includegraphics[width=0.8\textwidth]{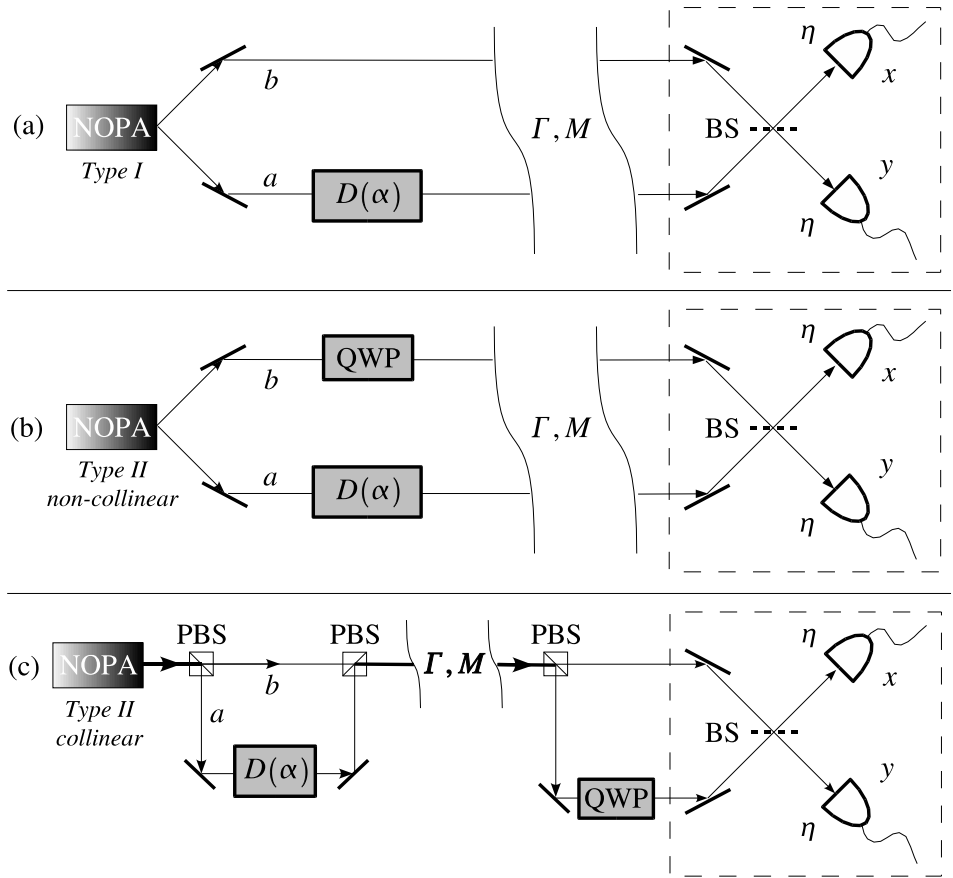}
\end{center}
\caption{Schematic diagrams of possible implementations of entanglement
based binary optical communication with TWB and multiport homodyne detection.
In (a): TWB are produced by spontaneous downconversion in a type I
nondegenerate optical parametric amplifier (NOPA). In this case the two 
modes have the same polarization. A displacement $D(\alpha)$ is applied to 
mode $a$ and then the signal is transmitted toward a multiport homodyne receiver. 
In (b): TWB are produced by spontaneous downconversion in a noncollinear 
type II NOPA. In this case the two TWB modes have different wave-vector 
and orthogonal polarizations. A displacement $D(\alpha)$ is applied to mode $a$,
while a quarter wave plate (QWP) rotates the polarization of the mode $b$
so that they become same polarized.  After the propagation, the
two modes are detected as before. 
In (c): TWB are produced by spontaneous downconversion in a collinear 
type II NOPA. In this case the two TWB modes have the same wave-vector and 
orthogonal polarizations. 
A first polarizer beam splitter (PBS) reflects the
mode $a$, which is displaced by $D(\alpha)$, and transmits mode $b$; the
two modes are then recombined in a second PBS and transmitted. Before
detection another PBS reflects mode $a$ toward a QWP
that changes its polarization to the same of mode $b$.
During propagation modes are subjected to losses (at rate $\Gamma$) and 
thermal noise (parameter $M$). The detector efficiency is $\eta$.
\label{f:het}}
\end{figure}

Binary optical communication assisted by entanglement may be implemented
using twin-beam (TWB) state of two modes of radiation \cite{par}. Schematic
diagrams of some possible implementations are given in figure \ref{f:het}.
In the Fock basis the TWB writes as follows
\begin{equation}\label{twb:state}
\dket{\lambda}=\sqrt{1-\lambda^2} \, \sum_{n} \, \lambda^n \,
\ket{n}\ket{n}\:,
\end{equation}
where $|\lambda |<1$ and, without loss of generality, it may be taken as
real ($\lambda$ is sometimes referred to as the TWB parameter).  TWB is the
maximally entangled state (for a given, finite, value of energy) of two
modes of radiation. It can be produced either by mixing two single-mode
squeezed vacuum (with orthogonal squeezing phases) in a balanced beam
splitter \cite{kim} or, from the vacuum, by spontaneous downconversion in a
nondegenerate optical parametric amplifier (NOPA) made either by type I or
type II second order nonlinear crystal \cite{kum0}. Referring to the
amplification case, the evolution operator reads as $U_r = \exp{\left\{r
\left(a^\dag b^\dag-ab\right)\right\}}$ where the ``gain'' $r$ is
proportional to the interaction-time, the nonlinear susceptibility, and the
pump intensity.  We have $\lambda=\tanh r$, whereas the number of photons
of TWB is given by $N_\lambda =2\sinh^2 r=2\lambda^2/(1-\lambda^2)$. 
\par
The two signals to be discriminated in a AMK encoding are given by
$\dket{\psi_0} = \dket{\lambda}$ and $\dket{\psi_{\alpha}} =
[D_a(\alpha)\otimes \mathbb{I}] \dket{\lambda}$, where the displacement
operator $D_a(\alpha)$ is acting on one of the modes, say $a$, of the TWB.
The energy of the TWB channel, {\em i.e.} the average photon number {\em
per use}, is given by $N =\frac12 {\rm Tr}\{ (a^{\dag}a\otimes\mathbb{I} +
\mathbb{I}\otimes b^{\dag}b)
(\rho_0+\rho_{\alpha})\}=N_{\lambda}+\frac12|\alpha|^2$, with $\rho_0 =
\dket{\psi_0}\dbra{\psi_0}$ and $\rho_{\alpha} =
\dket{\psi_{\alpha}}\dbra{\psi_{\alpha}}$.
\par 
The error probability for the ideal discrimination between the two
states $\dket{\psi_{0}}$ and $\dket{\psi_{\alpha}}$ reads as follows
\begin{eqnarray}\label{Q:het:id}
Q_{{\rm e}} &=& \frac{1-\sqrt{1 - |\dbraket{\psi_0}{\psi_{\alpha}}|^2}}{2}
\nonumber\\
\mbox{} &=& \frac{1-\sqrt{1-\exp\{ -2N(1-\beta)(1+\beta N)
\}}}{2},
\end{eqnarray}
where $\beta \equiv N_{\lambda}/N$ is the fraction of the channel energy 
that is used to establish the entanglement between the two modes.
Probability (\ref{Q:het:id}) is minimum for $\beta = (N-1)/2N$, when $N\ge
1$, and for $\beta=0$ when $0<N<1$. In summary one has
\begin{eqnarray}
Q_{{\rm e}} = P_{{\rm e}} &(N < 1) 
\label{en0} \\
Q_{{\rm e}} = \frac{1-\sqrt{1-\exp\{-\frac12(1+N)^2\}}}{2}\qquad &(N\ge 1) 
\label{en1}\:,
\end{eqnarray}
where $P_{\rm e}$, given in equation (\ref{optPe}), is the minimum error 
probability for single-mode AMK signals. Equations (\ref{en0}) and 
(\ref{en1}) say that $Q_{{\rm e}} \leq P_{{\rm e}}$ $\forall N$,
{\em i.e.} that the use of entanglement, at least in the ideal situation
considered so far, never increases the error probability, and it is
convenient if the photon number of the channel is larger than one.
\par
In order to see whether this result holds also in practice, it is necessary
to find out a realistic receiver able to discriminate
$|\psi_0\rangle\rangle$ and $|\psi_\alpha\rangle\rangle$, and to discuss
its performances in presence of noise. As concern detectors, we may use either
multiport homodyne detection, if the two modes have the same frequencies
\cite{wal}, or heterodyne detection otherwise \cite{yue}. Both these
detection schemes allow the measurement of the real and the imaginary part
of the complex operator $Z = a-b^{\dag}$.  In figure \ref{f:het} we have
referred to to eight-port homodyne detection; however all the results also
hold for other multiport homodyne schemes and for heterodyne detection.  Each
outcome from the measurement of $Z$ is a complex number $z$ and the POVM of
receiver is given by $\Pi_z = \frac{1}{\pi} \dket{z}\dbra{z}$, where
$$\dket{z}=[D_a(z)\otimes \mathbb{I}]\sum_n \ket{n}\ket{n} = [\mathbb{I}\otimes
D_b(-z^{*})]\sum_n \ket{n}\ket{n}\:.$$
\par
\begin{figure}[htb]
\begin{center}
\includegraphics[width=0.5\textwidth]{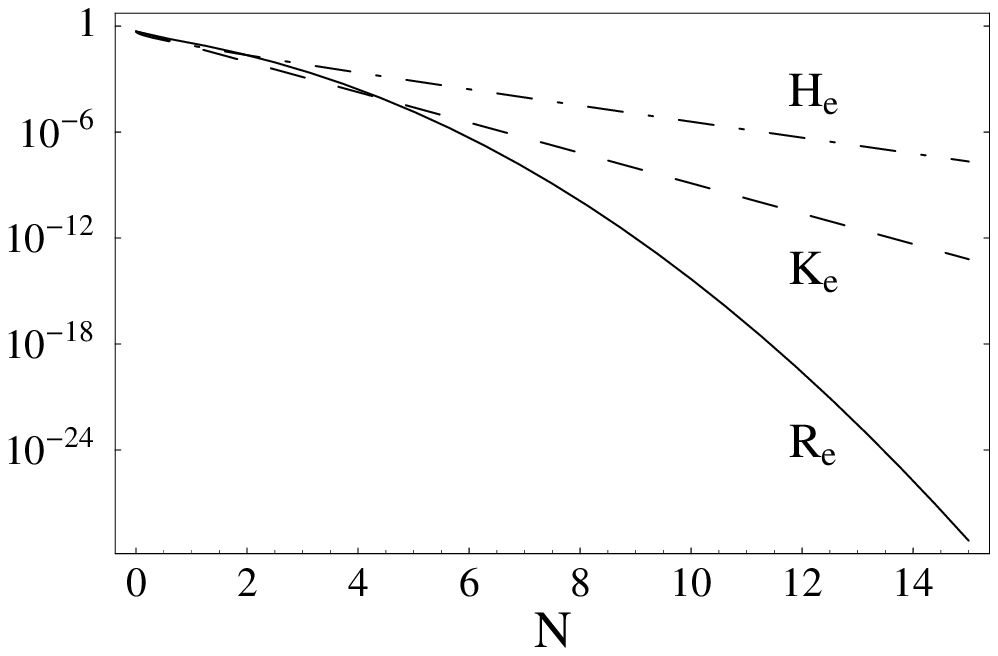}\\
\includegraphics[width=0.6\textwidth]{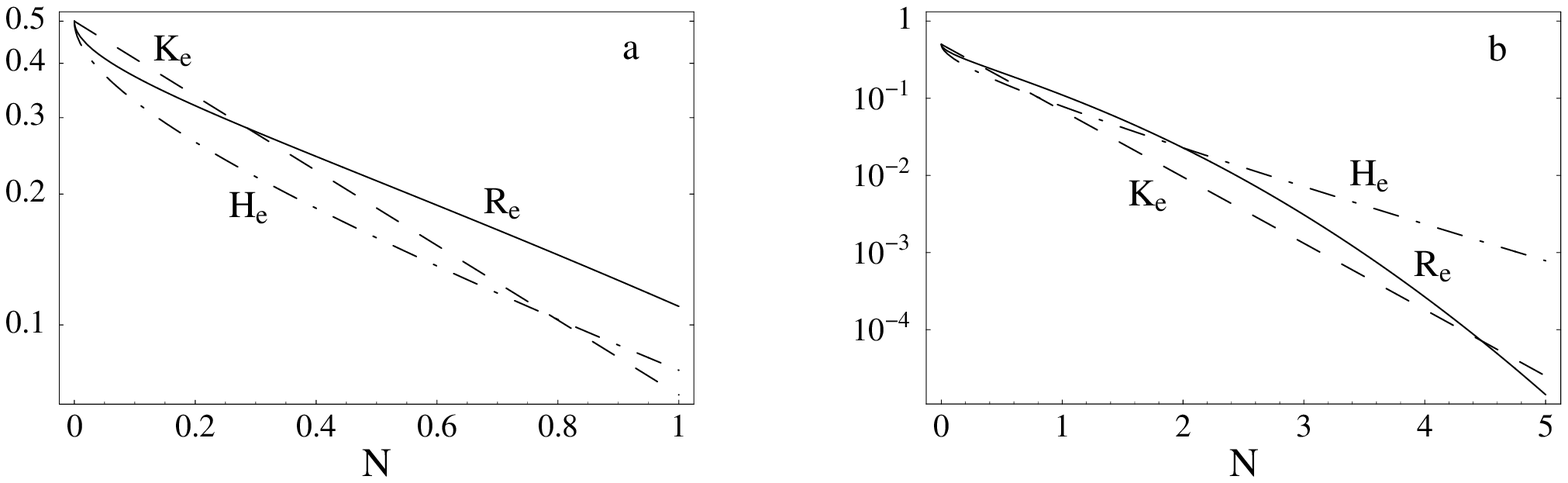}
\end{center}
\caption{Linear-log plot of the error probabilities of direct ($K_{\rm
e}$, dotted line), homodyne ($H_{\rm e}$, dot-dashed line) and heterodyne
($R_{\rm e}$, solid line) detection in the ideal case (absence of noise in
propagation and detection) as functions of the channel energy $N$. We chose
$\cos\phi^2 = 0.99$ for direct detection. As the channel energy
increases, the heterodyne error probability decreases more rapidly than the
single-mode ones.  In the smaller pictures: magnification of the plot in
the regions (a): $0<N<1$ and (b): $0<N<5$. For a channel energy $N\lesssim
0.79$ homodyne detection gives the best performances, whereas as the
channel energy increases the best results are obtained by the direct
($0.79\lesssim N \lesssim 4.46$) and heterodyne detection
($N\gtrsim 4.46$).  \label{f:khhvsid}}
\end{figure}

 
As already discussed for homodyne detection, we may take the amplitude
$\alpha$ as real. In this case a suitable inference rule to infer the
input state from $Z$-data involves the real part of the outcome as follows
$${\rm Re}[z]>\Lambda \Longrightarrow \dket{\psi_{\alpha}}\:,$$
where $\Lambda$ is a threshold value, which should be chosen such to
minimize the probability of error
\begin{equation}\label{err:het:id:a}
R_{{\rm e}}=\frac12 \left\{R(0|\alpha)+R(\alpha|0)\right\}\:,
\end{equation}
where $R(0|\alpha)$ and $R(\alpha|0)$ are the probabilities to detect
$\dket{\psi_{0}}$ when $\dket{\psi_{\alpha}}$ was sent and \emph{vice
versa}. The heterodyne distribution conditioned to a
displacement $D(\alpha)=D_a(\alpha)\otimes \mathbb{I}$ is given by the
probability density
\begin{eqnarray} \label{het:distrib:id}
r(z|\mu)=|\dbra{z} D(\mu)\dket{\lambda}|^2=\frac{1}{\pi \Delta_\lambda^2}
\exp\left\{ -\frac{|\mu-z|^2}{\Delta_{\lambda}^2} \right\}
\end{eqnarray}
with $\mu=0,\alpha$, and $\Delta_\lambda^2 = (1-\lambda) /
(1+\lambda)=(\sqrt{N_\lambda+2} -\sqrt{N_\lambda}) /
(\sqrt{N_\lambda+2}+\sqrt{N_\lambda})$. Therefore we have 
\begin{eqnarray}
R(0|\alpha)&=&\int_{-\infty}^{\Lambda} {\rm d}x \int_{-\infty}^{\infty}
{\rm d}y\, r(z|\alpha)\\
R(\alpha|0)&=&\int_{\Lambda}^{\infty} {\rm d}x \int_{-\infty}^{\infty} {\rm
d}y\, r(z|0),
\end{eqnarray}
where $z=x+i y$, and the error probability (\ref{err:het:id:a}) 
becomes
\begin{equation}\label{err:het:id:b}
R_{{\rm e}}=\frac12 \left\{ 1- \frac12 \left( {\rm
Erf}\left[\frac{\Lambda}{\Delta_{\lambda}} \right] + {\rm Erf}\left[
\frac{\alpha - \Lambda}{\Delta_{\lambda}} \right]\right) \right\}.
\end{equation}
$R_{\rm e}$ in equation (\ref{err:het:id:b}) is minimized by choosing
$\Lambda=\alpha/2$, thus leading to
$$R_{{\rm e}}=\frac12\left\{ 1- {\rm Erf}\left[\frac12
\frac{\alpha}{\Delta_{\lambda}}\right] \right\}\:.$$ 
At this point, the error probability $R_{{\rm e}}$ can be further minimized
by tuning the entanglement fraction $\beta$. By substituting the expression
for the amplitude $\alpha = \sqrt{2 N (1-\beta)}$ and the variance
$\Delta_{\lambda}^2 =(\sqrt{\beta N+2}-\sqrt{\beta N})/(\sqrt{\beta
N+2}+\sqrt{\beta N})$, we obtain
\begin{equation}
R_{{\rm e}}=\frac12 \left\{ 1-{\rm Erf}\left[ \frac12
\sqrt{\frac{2N(1-\beta)(\sqrt{\beta N+2}+\sqrt{\beta N})}{\sqrt{\beta
N+2}-\sqrt{\beta N}}} \right] \right\}.
\end{equation}
The optimal entanglement fraction, which minimizes $R_{{\rm e}}$, is given by
\begin{equation}\label{beta:opt:ideal}
\beta_{{\rm opt}}(N)=\frac{N}{2(1+N)}\,.
\end{equation}
In figure \ref{f:khhvsid} we report the resulting expression for the error
probability, compared with the corresponding single-mode error
probabilities $K_{\rm e}$ and $H_{\rm e}$. As the channel energy increases,
the heterodyne error probability decreases more rapidly than the
single-mode ones.  For a channel energy $N\lesssim 0.79$ homodyne detection
gives the best performances, whereas as the channel energy increases the
best results are obtained by direct detection ($0.79\lesssim N \lesssim
4.46$) and heterodyne detection ($N \gtrsim 4.46$). These results are
summarized in table \ref{t:uno}. Notice that for $N \gtrsim 5.2$ we have
$R_{\rm e} < P_{\rm e}$, {\em i.e.} TWB heterodyne channel provides better
performance even than ideal single-mode channel.
\subsection{Heterodyne detection - Noise in propagation and detection}
At first we consider the noise occurring during the propagation of the
TWB or its displaced version. This is described as the coupling of
each mode of the TWB with a thermal bath of oscillators at temperature $T$.
The dynamics is then described by the two-mode Master equation
\begin{eqnarray}\fl
\frac{{\rm d}\rho_t}{{\rm d}t} = \left\{ \Gamma (1+M) L[a]+\Gamma (1+M)
L[b]+\Gamma M L[a^{\dag}] + \Gamma M L[b^{\dag}]\right\}\rho_t \,,
\label{mast:eq:diss}
\end{eqnarray}
where $\rho_t$ is the density matrix of the bipartite system and the other
parameters are as in equation (\ref{mast:eq:diss:one:mode}).  The terms
proportional to $L[a]$ and $L[b]$ describe the losses, whereas the terms
proportional to $L[a^{\dag}]$ and $L[b^{\dag}]$ describe a linear
phase-insensitive amplification process. Of course, the dissipative
dynamics of the two modes are independent on each other.
\par
\begin{figure}[htb]
\begin{center}
\includegraphics[width=0.7\textwidth]{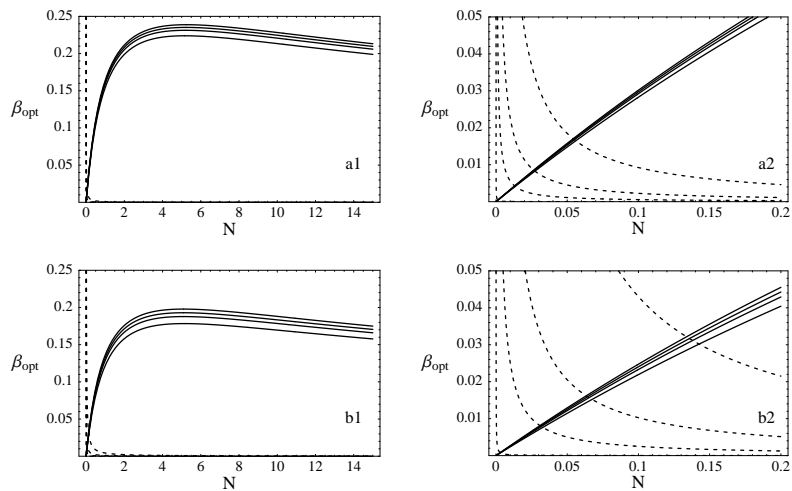}
\end{center}
\caption{Plots of the optimal entanglement fraction $\beta_{\rm opt}$
(solid lines) given in equation (\ref{beta:opt:all}) and of the survival 
entanglement fraction $\beta_{\rm s}$ (dashed lines) given in equation
(\ref{surv:beta}) as a function of the channel energy $N$ for different
values of $\Gamma$, $M$ and $\eta$. The plots (a2) and (b2) are magnifications
of the regions $0<N<0.2$ of plots (a1) and (b1), respectively. In all the 
plots we put (from top to bottom for solid lines, from bottom to top for
dashed lines) $M=5\,10^{-3}, 5\,10^{-2}, 0.1$ and $0.2$, respectively,
whereas: (a1) $\eta=0.9$, $\Gamma t=10^{-1}$; (b1) $\eta=0.9$, 
$\Gamma t=2\,10^{-1}$. As the channel becomes more noisy, entanglement
becomes less useful. Notice that for this choice of the parameters, when
$N\gtrsim 0.056$ (a2) and $N\gtrsim 0.142$ (b2) the state arriving at the
receiver is always non separable. \label{f:betavsent}}
\end{figure}

The Master equation (\ref{mast:eq:diss}) can be reduced to a
Fokker-Planck equation for the two-mode Wigner function
$W_{\mu}(\xi,\zeta)\equiv W[\rho_{\mu}](\xi,\zeta)$ of the system,
\begin{equation}\fl
W_{\mu}(\xi,\zeta) \equiv \frac{1}{\pi^4} \int\! {\rm d}^2 \chi \int\! {\rm
d}^2 \lambda \, {\rm e}^{\xi \chi^* -\xi^* \chi} \, {\rm e}^{\zeta
\lambda^* -\zeta^* \lambda} \, {\rm Tr}\left\{ \rho_{\mu} \,
D_{a}(\chi)\otimes D_{b}(\lambda)\right\}\,,
\end{equation}
where $\xi,\zeta \in \mathbb{C}$, $\rho_{\mu} =
\dket{\psi_{\mu}}\dbra{\psi_{\mu}}$, $\mu=0,\alpha$, and $D_{j}$, $j=a,b$, 
is the displacement operator acting on mode $j$. Using the differential
representation of the superoperator in equation (\ref{mast:eq:diss}), the
corresponding Fokker-Planck equation reads as follows
\begin{equation}\label{mast:eq:diss:wig}\fl
\partial_t W_{\mu,t}(\xi,\zeta) = \frac{\Gamma}{2}\left\{
\sum_{j=1}^{2}\left( \partial_{x_j}x_j +
\partial_{y_j}y_j\right) + \left( 2M+1 \right)
\sum_{j=1}^{2}\left( \partial_{x_j x_j}^{2} + \partial_{y_j y_j}^{2}\right)
\right\}W_{\mu,t}(\xi,\zeta)\,,
\end{equation}
where $\xi=x_1+i y_1$, $\zeta=x_2+i y_2$.
The solution of equation (\ref{mast:eq:diss:wig}) can be written as
\begin{equation}\label{wig:sol:mast}\fl
W_{\mu,t}(\xi,\zeta) = \int \!\!\! \int\! {\rm d}x'_1 {\rm d}y'_1 \int \!\!\!
\int\! {\rm d}x'_2 {\rm d}y'_2 \, W_{\mu,0}(x'_1, y'_1; x'_2, y'_2) \times
\prod_{j=1}^{2} G_{t}(x_j|x'_j)G_{\tau}(y_j|y'_j) \,,
\end{equation}
where $W_{\mu,0}(x'_1, y'_1; x'_2, y'_2)$ is the Wigner 
function at $t=0$ and the Green's functions $G_{t}(x_j|x'_j)$ are given in
equation (\ref{gre}). The Wigner function $W_{\mu,0}(x_1,y_1;x_2,y_2)$
before the propagation is given by (remind that $\mu=0,\alpha$, and
$\alpha$ is taken as real)
\begin{eqnarray}\label{twb:displ:wig}\fl
W_{\mu,0} = \frac{\exp\left\{ -\frac{(x_1+x_2-\mu)^2}{4
\sigma_{+}^2} -\frac{(y_1+y_2)^2}{4 \sigma_{-}^2}
-\frac{(x_1-x_2-\mu)^2}{4
\sigma_{-}^2} -\frac{(y_1-y_2)^2}{4
\sigma_{+}^2}\right\}}{(2\pi\sigma_{+}^2)(2\pi\sigma_{-}^2)}\:
\end{eqnarray}
with
\begin{equation}
\sigma_{\pm}^2 = \frac14 (\Delta_{\lambda}^2)^{\mp 1}\:. 
\end{equation}
$\Delta_{\lambda}^2$ being as in equation (\ref{het:distrib:id}). Since
$W_{\alpha,0}$ is Gaussian, the Wigner function $W_{\alpha,t}$ can be
easily evaluated. One has
\begin{eqnarray}\fl
W_{\mu,t} = \frac{\exp\left\{
-\frac{(x_1+x_2-\mu\: {\rm e}^{-\frac{1}{2} \Gamma t})^2}{4 \Sigma_{+}^2}
-\frac{(y_1+y_2)^2}{4 \Sigma_{-}^2} -\frac{(x_1-x_2-\mu\:
{\rm e}^{-\frac{1}{2}\Gamma t})^2}{4 \Sigma_{-}^2} -\frac{(y_1-y_2)^2}{4
\Sigma_{+}^2}\right\}}{(2\pi\Sigma_{+}^2)(2\pi\Sigma_{-}^2)} \,\label{Wevol}
\end{eqnarray}
where
\begin{equation}
\Sigma_{\pm}^2 = D^2 + \sigma_{\pm}^2 {\rm e}^{-\Gamma t}\,.
\end{equation}
The signals described by the Wigner functions of equation (\ref{Wevol})
correspond to entangled states if $\Sigma_-^2\leq \frac14$
\cite{paropt,geza,simon}, {\em i.e.} if
$$
{\rm e}^{-2r} \equiv 1+N_\lambda 
-\sqrt{N_\lambda(N_\lambda+2)} \leq
{\rm e}^{\Gamma t}-(2M+1)({\rm e}^{\Gamma t}-1)\,.
$$
If no thermal noise occurs ($M=0$), entanglement is present at any time,
whereas for $M\neq 0$ the survival time is given by 
\begin{eqnarray} 
t_{\rm s}= \frac1{\Gamma} \log\left(1+\frac{\sqrt{N_\lambda(N_\lambda+2)}
-N_\lambda}{2M}\right) \label{thre}\;,
\end{eqnarray}
and the corresponding survival entanglement fraction $\beta_{\rm s}(N,\Gamma,
M, t)$ reads as follows
\begin{eqnarray}
\beta_{\rm s}(N,\Gamma,M,t) = \frac{2 M^2 ({\rm e}^{\Gamma t}-1)^2}{N [1 -
2N({\rm e}^{\Gamma t}-1)]}
\label{surv:beta}\;.
\end{eqnarray}
The meaning of equation (\ref{surv:beta}) is that if the initial
entanglement fraction is above threshold, {\em i.e.} $\beta > \beta_{\rm
s}$, the state remains not separable after propagation.  Besides
propagation noise, one should take into account detection efficiency at the
heterodyne receiver.  In this case the POVM of the detector is a Gaussian
convolution of the ideal POVM
\begin{equation}\label{imperf:het}
\Pi_z(\eta)=\frac{1}{\pi \sigma_{\eta}^2} \int\! {\rm d}^2 \beta \,
\exp\left\{ - \frac{|z - \beta|^2}{\sigma_{\eta}^2} \right\}\,
\Pi_{\beta} \, ,
\end{equation}
with $\sigma_{\eta}^2=(1-\eta)/\eta$.  The Wigner function associated to
the POVM (\ref{imperf:het}) is given by 
\begin{equation}\fl
W[\Pi_{z}(\eta)](x_1,y_1;x_2,y_2)=\frac{\exp\left\{ -
\frac{(x_1-x_2-\hbox{Re}[z])^2}{\sigma_{\eta}^2} 
-\frac{(y_1+y_2-\hbox{Im}[z])^2}{\sigma_{\eta}^2}\right\}
}{\pi
\sigma_{\eta}^2} \end{equation}
and the corresponding heterodyne distribution $r_{\eta,\Gamma,M}(z|\alpha)$
by
\begin{eqnarray}
r_{\eta,\Gamma,M}(z|\alpha) = \frac{1}{\pi
\Delta_{\eta,\Gamma,M}^2} \exp\left\{ -\frac{|z-\alpha\: {\rm e}^{-\frac12
\Gamma t}|^2}{\Delta_{\eta,\Gamma,M}^2} \right\}\,,
\end{eqnarray}
with
\begin{eqnarray}
\Delta_{\eta,\Gamma,M}^2 =
4\Sigma_{-}^{2}+\sigma_{\eta}^2
= (1+2M)(1-{\rm e}^{-\Gamma t})+\Delta_{\lambda}^2\, {\rm e}^{-\Gamma t} +
\frac{1-\eta}{\eta}\,.
\end{eqnarray}
The error probability, as defined in the previous Sections,
already taking into account that the optimal threshold is given by 
$\Lambda=\frac12 \alpha\: {\rm e}^{-\frac12 \Gamma t}$, reads as follows
\begin{equation}
R_{{\rm e}}(N,\eta, \Gamma, M)=\frac12 \left\{ 1- {\rm Erf}\left[ \frac12
\frac{\alpha\: {\rm e}^{-\frac12 \Gamma t}}{\Delta_{\eta,\Gamma,M}} \right]
\right\}\,,
\end{equation}
and the optimal entanglement fraction $\beta_{{\rm opt}}(N,\eta,\Gamma,M,t)
\equiv \beta_{{\rm opt}}$, which minimizes the error probability, is
obtained after some algebra. One has
\begin{equation}\label{beta:opt:all}\fl
\beta_{{\rm opt}}=\frac{\eta^2 N {\rm e}^{-2\Gamma t}}{1+ A(N,\eta,\Gamma,M,t)
+ B(N,\eta,\Gamma,M) \sqrt{1+C(N,\eta,\Gamma,M,t)}}\,,
\end{equation}
where
\begin{eqnarray}
A(N,\eta,\Gamma,M)&=&\eta N {\rm e}^{-\Gamma t} (2 - \eta\: {\rm
e}^{-\Gamma t}) + 4 f(N,\eta,\Gamma,M)\,, \\
B(N,\eta,\Gamma,M)&=&1 + \eta N {\rm e}^{-\Gamma t} + 2 \eta M (1-{\rm
e}^{-\Gamma t})\,, \\
C(N,\eta,\Gamma,M)&=&2\eta N {\rm e}^{-\Gamma t}(1-\eta\: {\rm e}^{-\Gamma
t}) + 2 f(N,\eta,\Gamma,M)\,,
\end{eqnarray}
with
\begin{equation}
f(N,\eta,\Gamma,M) = \eta M(1 - {\rm e}^{-\Gamma t})[1 + \eta N {\rm
e}^{-\Gamma t}+ \eta M (1- {\rm e}^{-\Gamma t})]\,.
\end{equation}
As a matter of fact, the optimal entanglement fraction
depends on the noise parameters of the channel. In particular, 
it can be seen that $\beta_{{\rm opt}}$ decreases 
as $\Gamma$ and $M$ increase and $\eta$ decreases. In other words, 
as the channel becomes more noisy, the entanglement becomes less useful.
In figure \ref{f:betavsent} we report $\beta_{{\rm opt}}$ for fixed
detection efficiency and $\beta_{s}$ as functions of the 
channel number of photons for different values of the other parameters.
\subsection{Direct vs heterodyne detection}
\begin{figure}[htb]
\begin{center}
\includegraphics[width=0.7\textwidth]{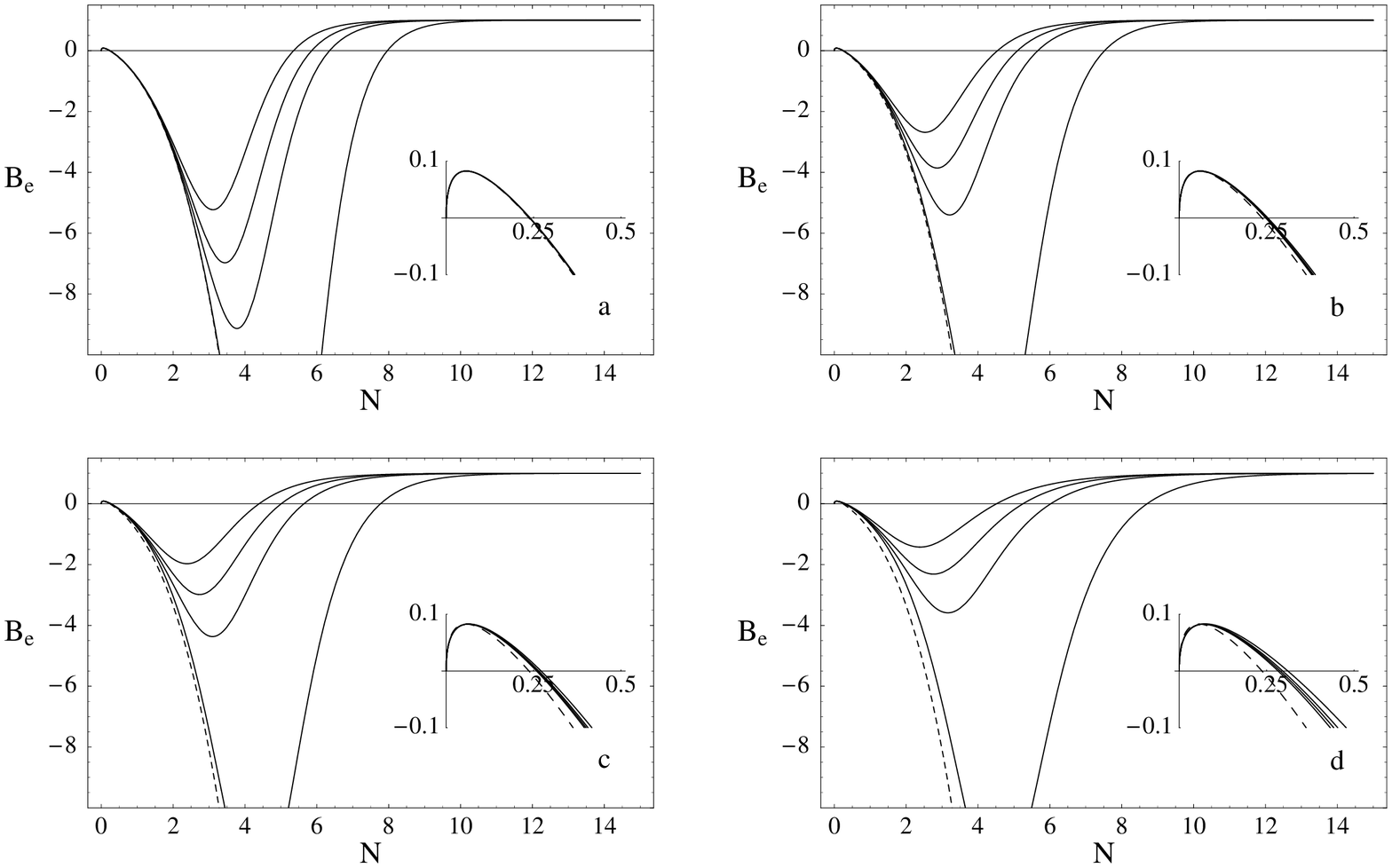}
\end{center}
\caption{Plots of $B_{{\rm e}}(N,\eta_{{\rm ken}},\eta_{{\rm het}},\Gamma,
M)$, defined in equation (\ref{B:e}), as a function of $N$ for different values
of $\Gamma t$ and $M$. The {\sc on/off} and heterodyne detection
efficiencies are chosen as $\eta_{{\rm ken}}=0.95$
and $\eta_{{\rm het}}=0.85$, respectively, and $\cos^2\phi=0.99$. The
dashed line is $B_{{\rm e}}$ with $\Gamma, M = 0$. In all the plots the
solid lines correspond to (from bottom to top) $M=5\, 10^{-3}, 5\, 10^{-2},
0.1$ and $0.2$, respectively, whereas $\Gamma t$ is: (a)  $\Gamma
t=10^{-2}$, (b) $\Gamma t=5\, 10^{-2}$, (c) $\Gamma t=0.1$, (d) $\Gamma
t=0.2$. The insets refer to the region $0<N<0.5$. When $B_{{\rm e}}>0$ one
has $R_{{\rm e}}<K_{{\rm e}}$.\label{f:kenvshet}}
\end{figure}

Heterodyne detection provides better performance than direct detection 
when the quantity
\begin{equation}\label{B:e}
B_{{\rm e}}(N,\eta_{{\rm ken}},\eta_{{\rm het}},\Gamma, M) = 1 -
\frac{R_{{\rm e}}(N,\eta_{{\rm het}},\Gamma,M)}{K_{{\rm e}}(N,\eta_{{\rm
ken}},\Gamma,M)}\,,
\end{equation}
is positive, {\em i.e.} $R_{\rm e}<K_{\rm e}$. In equation (\ref{B:e})
$\eta_{\rm ken}$ and $\eta_{\rm het}$ are the {\sc on/off} and heterodyne
detection efficiencies, respectively. In figure \ref{f:kenvshet} we plot
$B_{\rm e}$ as a function of channel energy $N$ for different values of the
other parameters. Three regions of interest can be identified, since, as in
the case of homodyne detection, for given values of $\eta_{{\rm ken}}$,
$\eta_{{\rm het}}$, $\Gamma$ and $M$ we have two different thresholds for
the energy channel, namely $N^{\rm (het)}_{{\rm th},j}(\eta_{{\rm
ken}},\eta_{{\rm het}},\Gamma,M)$, $j=1,2$, such that $B_{{\rm e}}=0$.
Heterodyne error probability is the smallest in a small region for $N \le
N^{\rm (het)}_{{\rm th},1}$ and for large values of the channel energy ($N
> N^{\rm (het)}_{{\rm th},2}$), whereas in an intermediate interval of
energy values ($N^{\rm (het)}_{{\rm th},1} < N \le N^{\rm (het)}_{{\rm
th},2}$) the best results are obtained by direct detection (see table
\ref{t:tre}).
\par
As for the homodyne detection, heterodyne receiver provides better results
for either small or large values of the channel energy $N$, whereas
for the intermediate region, whose width depends on $\Gamma$, $M$ and
$\eta$ as in the case of $A_{\rm e}$, direct detection should be preferred. 
\subsection{Heterodyne vs homodyne detection} 
\begin{figure}
\begin{center}
\includegraphics[width=0.7\textwidth]{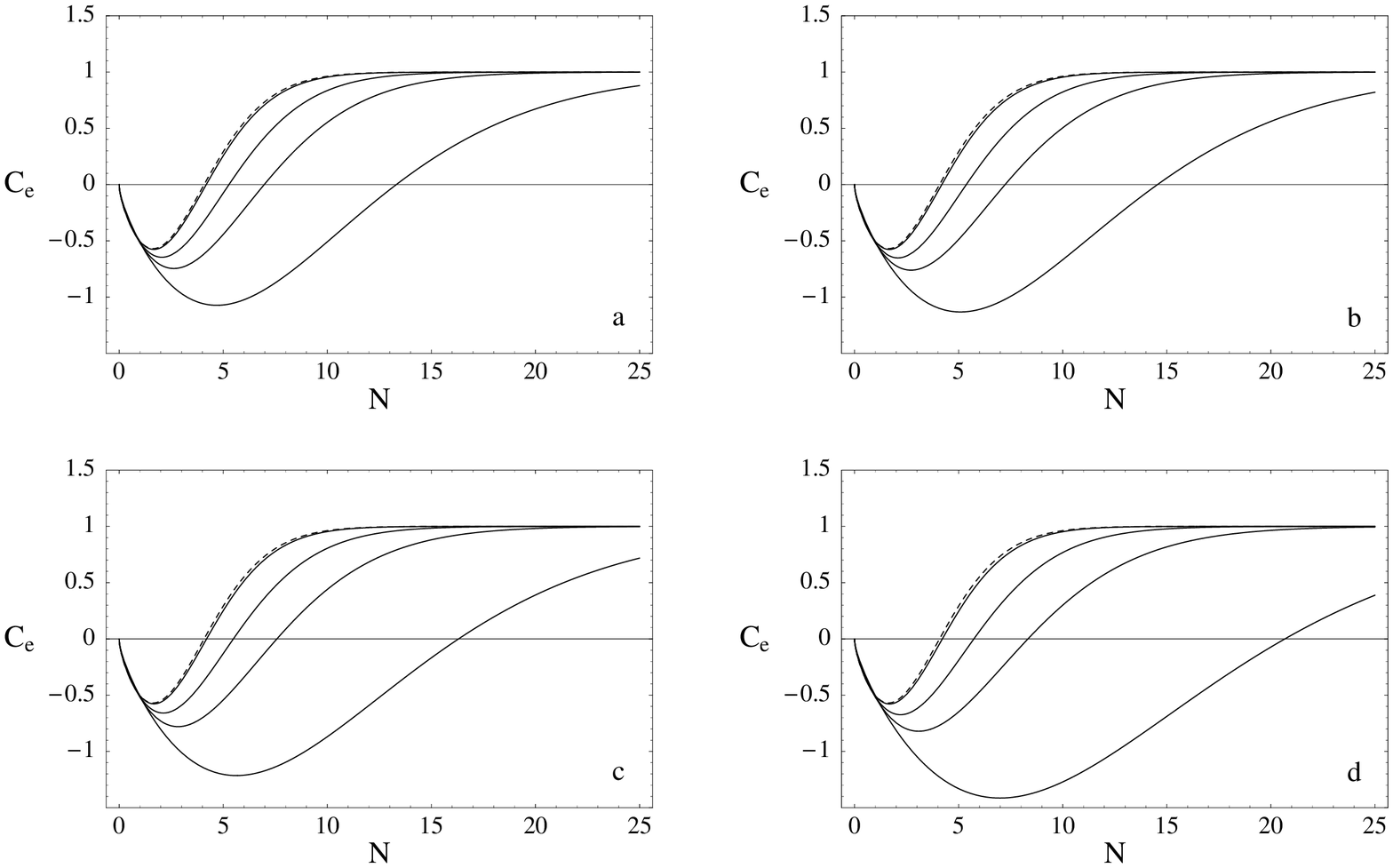}
\end{center}
\caption{Plots of $C_{{\rm e}}(N,\eta,\Gamma, M)$, defined in equation
(\ref{C:e}), as a function of $N$ for different values of $\Gamma t$ and
$M$. The heterodyne and homodyne detection efficiencies are chosen as 
$\eta_=0.85$. The dashed line is $C_{{\rm e}}$ with
$\Gamma, M = 0$. In all the plots the solid lines correspond to (from top
to bottom) $M=5\, 10^{-3}, 5\, 10^{-2}, 0.1$ and $0.2$, respectively,
whereas $\Gamma t$ is: (a)  $\Gamma t=10^{-2}$, (b) $\Gamma t=5\, 10^{-2}$,
(c) $\Gamma t=0.1$, (d) $\Gamma t=0.2$. When $C_{{\rm e}}>0$ one has
$R_{{\rm e}}<H_{{\rm e}}$.\label{f:hetvshom}}
\end{figure}

\begin{figure}[htb]\begin{center}
\includegraphics[width=0.7\textwidth]{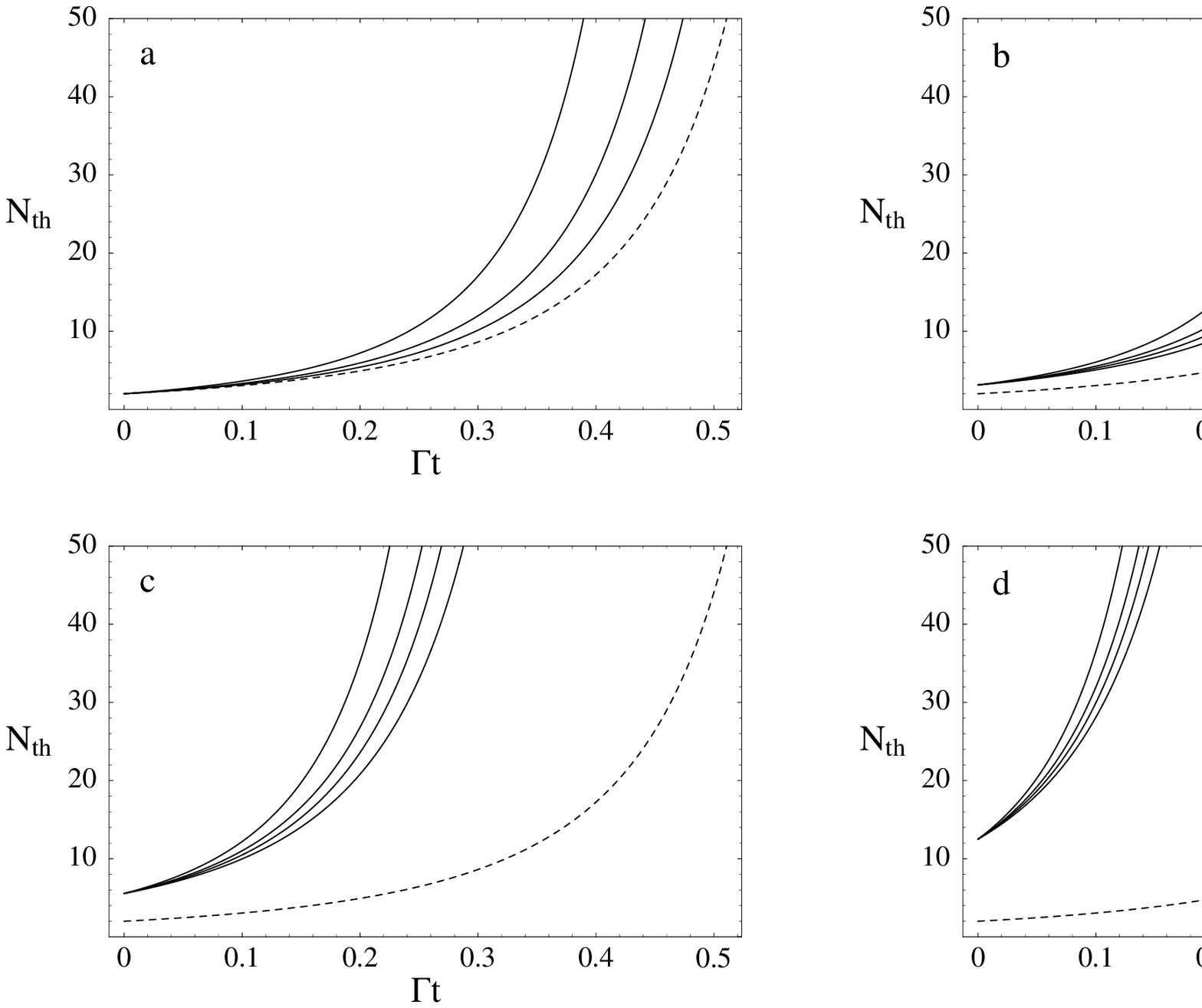}\end{center}
\caption{$N_{{\rm th}}$ as a function of $\Gamma t$ for different values of
the detection efficiency $\eta$ and the average number of thermal photons
$M$. When $N>N_{{\rm th}}$ the error probability using entangled channels and
heterodyne detection is less than the one obtained with a single-mode
homodyne detection ($R_{{\rm e}}<H_{{\rm e}}$).  The dashed line is
$N_{{\rm th}}$ for $\eta=1$ and $M=0$, while the solid lines represent the
threshold for (from right to left) $M=0, 5\, 10^{-2}, 0.1$ and
$0.2$, respectively, whereas: (a) $\eta=1$, (b) $\eta=0.9$, (c) $\eta=0.8$
and (d) $\eta=0.7$.}\label{f:figeta}
\end{figure}

Since the channel energy intervals where heterodyne and homodyne
detection should be preferred to direct one are quite similar, 
it is useful to introduce the function
\begin{equation}\label{C:e}
C_{{\rm e}}(N,\eta,\eta,\Gamma, M) = 1 -
\frac{R_{{\rm e}}(N,\eta,\Gamma,M)}{H_{{\rm e}}(N,\eta,\Gamma,M)}\,,
\end{equation}
which is positive when $R_{\rm e} < H_{\rm e}$. As one can see in figure
\ref{f:hetvshom}, there exists a threshold $N_{{\rm th}}(\eta,\Gamma,M)$ on
the channel energy $N$ such that if $N > N_{{\rm th}}$ then $R_{\rm e} <
H_{\rm e}$ (see table \ref{t:quattro}). Notice that the homodyne and
heterodyne detection efficiencies have the same value. In figure
\ref{f:figeta} we plot $N_{{\rm th}}$ for different physical situation. The
threshold increases with increasing noise either in the propagation or in
the detection stage.  For a channel energy above the threshold the use of
entanglement improves the communication performances.
\section{Conclusions}\label{Sec:Concl}
In this paper we have analyzed binary communication in single-mode 
and entangled quantum noisy channels. We took into account different
kind of noise that may occur, {\em i.e.} losses and 
thermal noise during propagation, non unit quantum efficiency
of detectors during the measurement stage.
\par
As concern single mode communication, we found that, in presence of noise,
homodyne detection is a more robust receiver when compared to direct detection. 
In particular, homodyne detection achieves a smaller error
probability for either small or large values of the energy of the channel, 
whereas for intermediate values direct detection should be preferred. 
\par
We then considered an entanglement based quantum channel build by amplitude
modulated twin-beam and multiport homodyne detection. As for the
homodyning, heterodyne detection should be preferred to direct detection
for either small or large values of the energy of the channels.  On the
other hand the comparison between the performances of heterodyne and
homodyne detection shows that there exists a threshold on the channel
energy, above which the error probability using entangled channels and
heterodyning  is the smallest. The threshold depends on the amount of
noise, and increases as imperfections in propagation and detection become
more relevant. We summarized our results in tables \ref{t:uno}, \ref{t:due},
\ref{t:tre}, and \ref{t:quattro}.
\par
We conclude that entanglement is a useful resource to improve binary
communication in presence of noise, especially in the large energy regime.  
\section*{Acknowledgments}
This work has been supported by the INFM project PRA-2002-CLON and by MIUR
through the PRIN project {\em Decoherence control in quantum information
processing}.  MGAP is research fellow at {\em Collegio Alessandro Volta}.

\vfill
\begin{table}[h]
\begin{center}
\begin{tabular}{c|c}
\hline\hline
{\bf Channel energy} & {\bf Best detector (channel)} \\
\hline\hline
$N \lesssim 0.79$ & \multicolumn{1}{l}{Homodyne (single mode)}\\
\hline
$0.79 \lesssim N \lesssim 4.46$ &
\multicolumn{1}{l}{Direct (single mode)} \\
\hline
$N \gtrsim 4.46$ & \multicolumn{1}{l}{Heterodyne (entangled)} \\
\hline
\end{tabular}
\end{center}
\caption{Comparison among direct, homodyne and heterodyne detection
as a function of the channel energy when transmission and detection are
ideal (see also figure \ref{f:khhvsid}).\label{t:uno}}
\end{table}
\begin{table}[h]
\begin{center}
\begin{tabular}{c|c}
\hline\hline
{\bf Channel energy} & {\bf Best detector (channel)} \\
\hline\hline
$N \leq N^{\rm (hom)}_{{\rm th}, 1}$ &
\multicolumn{1}{l}{Homodyne (single mode)}\\
\hline
$N^{\rm (hom)}_{{\rm th}, 1} < N \leq N^{\rm (hom)}_{{\rm th}, 2}$ &
\multicolumn{1}{l}{Direct (single mode)} \\
\hline
$N > N^{\rm (hom)}_{{\rm th}, 2}$ &
\multicolumn{1}{l}{Homodyne (single mode)} \\
\hline
\end{tabular}
\end{center}
\caption{Comparison between direct and homodyne detection as a function of the
channel energy in presence of noise during the transmission and detection
stages. The thresholds $N^{\rm (hom)}_{{\rm th}, j}(\eta_{\rm
ken},\eta_{\rm hom}, \Gamma,M)$, $j=1,2$, define three different regimes
(see also figure \ref{f:kenvshom}).\label{t:due}}
\end{table}
\begin{table}[h]
\begin{center}
\begin{tabular}{c|c}
\hline\hline
{\bf Channel energy} & {\bf Best detector (channel)} \\
\hline\hline
$N \leq N^{\rm (het)}_{{\rm th}, 1}$ &
\multicolumn{1}{l}{Heterodyne (entangled)}\\
\hline
$N^{\rm (het)}_{{\rm th}, 1} < N \leq N^{\rm (het)}_{{\rm th},2}$&
\multicolumn{1}{l}{Direct (single mode)} \\
\hline
$N > N^{\rm (het)}_{{\rm th}, 2}$ &
\multicolumn{1}{l}{Heterodyne (entangled)} \\
\hline
\end{tabular}
\end{center}
\caption{Comparison between direct and heterodyne detection as a function of the
channel energy in presence of noise during the transmission and detection
stages. The thresholds $N^{\rm (het)}_{{\rm th}, j}(\eta_{\rm
ken},\eta_{\rm het}, \Gamma,M)$, $j=1,2$, define three different regimes
(see also figure \ref{f:kenvshet}).\label{t:tre}}
\end{table}
\begin{table}[h]
\begin{center}
\begin{tabular}{c|c}
\hline\hline
{\bf Channel energy} & {\bf Best detector (channel)} \\
\hline\hline
$N \leq N_{{\rm th}}$ &
\multicolumn{1}{l}{Homodyne (single mode)}\\
\hline
$N > N_{{\rm th}}$&
\multicolumn{1}{l}{Heterodyne (entangled)} \\
\hline
\end{tabular}
\end{center}
\caption{Comparison between heterodyne and homodyne detection as a function
of the channel energy in presence of noise during the transmission and
detection stages. The threshold $N_{{\rm th}}(\eta,\Gamma,M)$ defines two
different regimes (see also figure \ref{f:hetvshom}).\label{t:quattro}}
\end{table}
\vfill

\begin{thebibliography}{99}
\bibitem{uir} This is a consequence of the fact that the set of 
displacement operators $\{D(\lambda)\}$, $\lambda \in \mathbb{C}$ represents
a unitary irreducible representation the Weyl-Heisenberg group.
\end{thebibliography}
\end{document}